\documentclass[5p]{elsarticle}
\usepackage{listings}
\usepackage{color}

\definecolor{myblue}{rgb}{0,0,0.9}
\definecolor{mygray}{rgb}{0.9,0.9,0.9}
\definecolor{mymauve}{rgb}{0.58,0,0.82}
\usepackage[cmex10]{amsmath}
\usepackage{graphicx}
\usepackage{epstopdf}
\journal{Journal of \LaTeX\ Templates}

\begin{document}

\begin{frontmatter}

\title{Future Large-Scale Memristive Device Crossbar Arrays: Limits Imposed by Sneak-Path Currents on Read Operations}


\author[rvt]{Yansong Gao\corref{cor1}}
\ead{yansong.gao@adelaide.edu.au}

\author[focal]{Omid Kavehei}
\ead{omid.kavehei@rmit.edu.au}
\author[els]{Damith C.~Ranasinghe}

%\ead[url]{http://www.elsevier.com}
\ead{damith@cs.adelaide.edu.au}
\author[rvt]{Said F.~Al-Sarawi}
\ead{said.alsarawi@adelaide.edu.au}
\author[rvt]{Derek Abbott}
\ead{derek.abbott@adelaide.edu.au}

\cortext[cor1]{Corresponding author}
\address[rvt]{School of Electrical and Electronic Engineering,
The University of Adelaide, Adelaide, SA, Australia 5005}
\address[focal]{School of Electrical and Computer Engineering, Royal Melbourne Institute of Technology, Victoria 3001, Australia}
\address[els]{School of Computer Science, The University of Adelaide, SA 5005, Australia}

%\author[mymainaddress]{Yansong Gao}
%\author[mysecondaryaddress]{Omid Kavehei}
%\author[mythirdaryaddress]{Damith C.Ranasinghe}
%\author[mymainaddress]{Said Al-Sarawi}
%\author[mymainaddress]{Derek Abbott}
%%%\ead[url]{www.elsevier.com}
%%%\author[mymainaddress,mysecondaryaddress]{Elsevier Inc}
%%%\ead[url]{www.elsevier.com}
%%
%%\author[mymainaddress]{School of Electrical and Electronic Engineering,\\
%%The University of Adelaide, Adelaide, SA, Australia 5000\corref{mycorrespondingauthor}}
%%\cortext[mycorrespondingauthor]{Corresponding author}
%%\ead{support@elsevier.com}
%%
%\address[mysecondaryaddress]{School of Electrical and Computer Engineering,\\Royal Melbourne Institute of Technology Victoria 3001, Australia}
%\ead{omid.kavehei@rmit.edu.au}
%\address[mythirdaryaddress]{School of Computer Science,\\
%The University of Adelaide, SA 5005, Australia}
%\ead{damith@cs.adelaide.edu.au}

\begin{abstract}
%Passive crossbar array is in principle the simplest functional electrical circuit. 
Passive crossbar arrays based upon memristive devices, at crosspoints, hold great promise for the future high-density and non-volatile memories. The most significant challenge facing memristive device based crossbars today is the problem of sneak-path currents. In this paper, we investigate a memristive device with intrinsic rectification behavior to suppress the sneak-path currents in crossbar arrays. The device model is implemented in Verilog-A language and is simulated to match device characteristics readily available in the literature. Then, we systematically evaluate the read operation performance of large-scale crossbar arrays utilizing our proposed model in terms of read margin and power consumption while considering different crossbar sizes, interconnect resistance values, HRS/LRS (High Resistance State/Low Resistance State) values, rectification ratios and different read-schemes. The outcomes of this study are understanding the trade-offs among read margin, power consumption, read-schemes and most importantly providing a guideline for circuit designers to improve the performance of a memory based crossbar structure.  In addition, read operation performance comparison of the intrinsic rectifying memristive device model with other memristive device models are studied. 
\end{abstract}


\end{frontmatter}

\section{Introduction}\label{Introduction}
The two-terminal memristive device based crossbar array is a promising candidate for future non-volatile memories attributing to its low-power and small-area overhead, 3D integration, fast switching speed, compatibility with standard CMOS technology and simple fabrication process~\cite{yang2013memristive,eshraghian2011memristor}. However, ultra-high density integration of memristive devices in crossbar arrays still faces a number of challenges such as magnitude of sneak-path (leakage) currents, relatively high variation of device operation, and poor reproducibility. The most significant of these challenges that must be overcome to realize large-scale crossbar array memory is that of sneak-path currents, which demands reduction in leakage currents through unselected devices while reading the memory array. 

%This requires utilization of a series element with the memristive device, known as selector device. The selector device can either be integrated and fabricated with each memristive device or comes as a device that is connected in series with the memristive device (e.g. an addressing transistor). Some memristive devices also show an intrinsic high-nonlinearity such as high self-rectification ratio, which in practice limits the sneak-path currents, and hence, effectively acts as a much needed cross-point cell element without a selector. 

\begin{figure}
\centering
\includegraphics[trim=0 0 0 0,clip,width=0.45\textwidth]{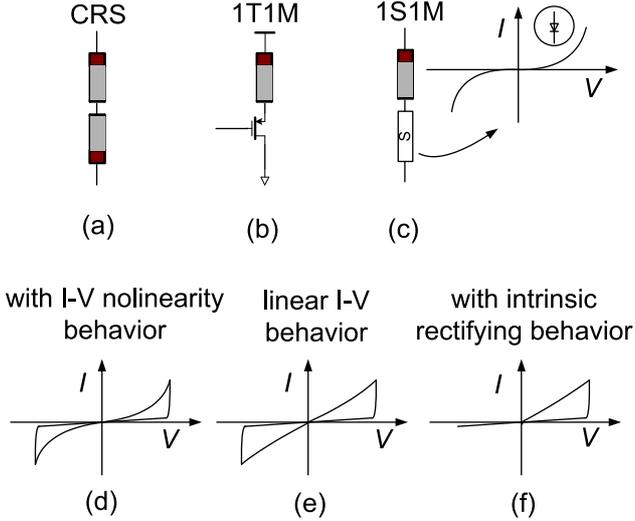}
\caption{I-V characteristic of selector.}
\label{IVCharac}
\end{figure}

To circumvent sneak-path currents in emerging crossbar memory arrays, several approaches have been proposed. One of those techniques uses two back-to-back bipolar memristive elements shown in Fig.\ref{IVCharac} (a). This approach resonates with the immense success of CMOS technology in which either the pull-up (composed of p-type transistor) or the pull-down (composed of n-type transistor) network is active at a given time; such a complimentary configuration greatly limits the short-circuit currents through the operating circuit. For memristive circuits a similar configuration, known as complementary resistive switch (CRS) is utilized~\cite{linn2010complementary,yang2012complementary,kavehei2013associative}, however, this configuration relies on the polarity difference between the two back-to-back connected memristive elements. Therefore, at a given time, only one of the devices is either in the LRS (Low Resistance State) or HRS (High Resistance State). This indicates that the total resistance of this configuration in equilibrium is always at the highest value, hence, limiting sneak-path currents through unselected cells within the crossbar array. Despite these features, this approach is shown to be not very successful given the relatively large footprint and significant hurdles in a successful miniaturization~\cite{yang2013memristive}. In addition, a reading operation on a CRS is destructive and the multilevel bits offered by a memristive device can no longer be exploited~\cite{zidan2013memristor}. 

Alternatively, the other two approaches using a selector device to reduce the sneak-path currents are: one-transistor-one-memristive device (1T1M), shown in Fig.~\ref{IVCharac} (b); and one-diode-one-memristive device (1D1M) structures~\cite{srinivasan2012punchthrough,zidan2013memristor}, shown in Fig.~\ref{IVCharac} (c). Using a transistor as the selecting device at each crosspoint raises the footprint of each cell and consequently its three-dimensional stacking capability. Therefore the 1T1M structure is unlikely to realize the potential scalability of memristive devices. In terms of the 1D1M structure, it suffers programming difficulty due to the significant voltage drop across the reverse-biased diode instead of the memristive device. Although a plausible solution is to use a special diode showing typical I-V (Current-Voltage) characteristic in Fig.~\ref{IVCharac} (c) to permit enough current density to write the serially connected memristive device by applying a higher voltage than threshold voltage, enabling the two terminal selector to show exponential current behavior~\cite{srinivasan2012punchthrough,lee2012varistor,kawahara20138}. However, the read operation performance does not always benefit from the nonlinear I-V characteristic, because of the voltage divider effect caused by the external selector~\cite{zhou2014crossbar}. In fact, the higher nonlinearity can diminish read operation performance if the nonlinearity is higher than an optimal value as will be discussed in Section~\ref{Compare1S1R}.

The other option is to engineer a large I-V nonlinearity as shown in Fig.~\ref{IVCharac} (d) by integrating an oxide layer to suppress about two orders of current magnitude smaller than that of the linear device---see Fig.~\ref{IVCharac} (e)~\cite{yang2012engineering}. However, to realize a large OFF/ON resistance window in this case, this kind of memristive device suffers from life-cycle limitations~\cite{yang2013memristive}.  

The memristive device with intrinsic rectifying behavior shown in Fig.~\ref{IVCharac} (f) has been empirically demonstrated  ~\cite{kim2010nanoscale,kim2011functional,lu2011two,tran2012self,lu2007non}, and a rectification ratio in the order of $ 10^6 $ has been achieved in \cite{kim2010nanoscale} without significant reduction in either switching speed or retention time, which illustrated its possible application as a solution to circumvent sneak-path currents. However, to the best of our knowledge, a detailed read operation performance evaluations of large-scale crossbar arrays based on memristive devices with intrinsic rectifying behavior has not been investigated in the literature. Therefore, our motivation in this paper is to systematically study the read operation performance of crossbar arrays using the memristive device with intrinsic rectifying behavior. In particular, we: (i) provide a Verilog-A behavioral model of a memristive device with intrinsic rectifying behavior based on published fabricated device parameters; (ii) evaluate read operation performance of large-scale arrays while considering different interconnect resistance values, HRS/LRS values, rectification ratios and different read-schemes utilizing our proposed model; (iii) demonstrate the effectiveness of this special type of memristive device as an approach to reduce sneak path effects in a crossbar structure; (iv) demonstrate the advantages of the memristive device with intrinsic-rectifying behavior by comparing with a linear memristive device (without rectification capability) and 1S1R structure, shown in Fig.~\ref{IVCharac} (e) and (c), respectively.
%The other potential solution is engineering nonlinearity into LRS state of memristive device~\cite{yang2012engineering} or integrating intrinsic diode rectifying behavior into memristive device~\cite{kim2010nanoscale,kim2011functional,lu2011two,tran2012self,lu2007non}. Although engineering nonlinearity into LRS state is one good solution, however, the high nonlinearity in ON state is hard to achieve~\cite{yang2012engineering}. Fortunately, as for memristive device with intrinsic rectification behavior, rectification ratio up to $ 10^6 $ has been demonstrated in \cite{kim2010nanoscale} without obvious compromise regarding to switching speed, retention time.
 
The rest of the paper is organized as follows: in Section~ \ref{CrossbarAndModel}, we describe read margin and read-schemes for a crossbar array. The memristive device behavioral model along with Verilog-A code are provided; in Section \ref{ResultsAndAnalysis} we evaluate our memristive device model within a passive crossbar array and analyze the read operation performance of crossbar arrays, specifically in terms of read margin and power consumption; read operation comparison with other memristive devices and discussions are presented in Section \ref{Comparison}; followed by conclusions in Section \ref{Conclu}.

\section{Crossbar Array and Device Model}\label{CrossbarAndModel}
\subsection{Crossbar Array}
\begin{figure}
\centering
\includegraphics[trim=0 0 0 0,clip,width=0.45\textwidth]{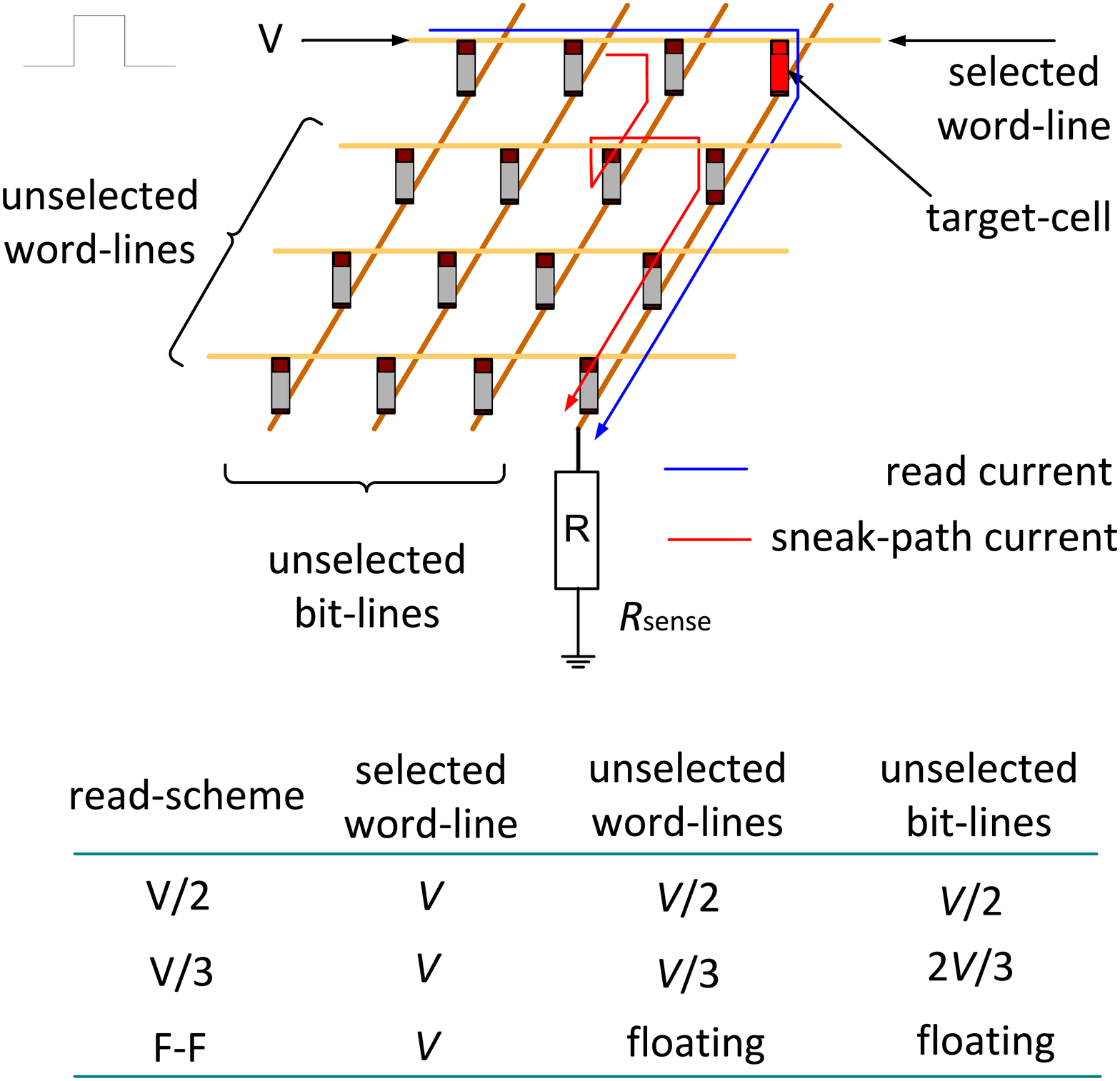}
\caption{Schematic of crossbar memory architecture. Here we define three different read-schemes according to the biased voltage on selected word/bit-line and unselected word/bit-lines.}
\label{crossbar}
\end{figure}

A crossbar array is comprised of two layers of parallel electrodes that are crossed perpendicularly, which act as word-lines or bit-lines. At each crosspoint, an element is formed, which can be programmed to the LRS or HRS to represent either a logic `1' or `0' when proper voltages are applied to word-lines and bit-lines. Unfortunately, as stated in Section~\ref{Introduction}, this promising memory architecture severely suffers from sneak-path currents, hence the need for appropriate approaches to suppress them. Because the sneak-path currents aggravate read/write operation performance of crossbar array and hence, limit the maximum size of a crossbar. The source of sneak-path current is shown in Fig.~\ref{crossbar}. As discussed earlier, one possible approach to suppress sneak-path currents and hence improve read/write operation performance is through using a memristive device with intrinsic rectifying behavior.

To simplify the read operation with large crossbars, in the following simulations a single sense resistor is used to sense the state of the memristive device instead of using sense amplifier. To obtain the best read margin, the sense resistor  value, $R_{\rm{sense}}$,  is calculated based on~\cite{zhou2014crossbar}:
 \begin{equation}\label{Rsense}
 R_{\rm sense}=\sqrt {R_{\rm ON}R_{\rm OFF}}.
 \end{equation}
The read margin, RM, is defined as:
 \begin{equation}
 \rm RM = {{\it V}_{\rm out}(\rm LRS) - {\it V}_{\rm out}(\rm HRS) \over {\it V}_{\rm WS}}
 \end{equation}
 where $ V_{\rm WS}  $ is the read voltage applied to the selected word-line shown in Fig.~\ref{crossbar}. While $ V_{\rm out}(\rm LRS) $ and $ V_{\rm out}(\rm HRS) $ are the voltages measured in $ R_{\rm sense} $ resistor when the target-cell is in LRS and HRS, respectively. In the literature, there are a number of read-schemes for crossbars that includes V/2, V/3 and F-F (Floating-Floating), where $V$ is the rail-to-rail (maximum) potential difference. The voltage bias requirements for these read-schemes are shown in Fig.~\ref{crossbar}.
 
Due to the influence from sneak-path currents and total interconnect resistance, the output voltage swing is dependent on the stored data and the position of the cell to be read within the array. Additionally, taking the interconnect resistance into account, the read margin will deteriorate further if the target-cell is located in the furthest corner from the word/bit-line voltage sources---in case of Fig.~\ref{crossbar}, this would be the top right corner cell. In this paper, we only evaluate operation performance under this worst-case position. 
\subsection{Memristive Device Model with Intrinsic Rectifying Behavior}
 The memristive device model with intrinsic rectifying behavior is investigated in this subsection. Besides the intrinsic rectifying behavior, the memristive device also shows well-defined threshold voltage and abrupt resistance switching. In the following, we use a simplified mathematical model for this special memristive device to match the rectifying and threshold voltage behaviors that are empirically reported in~\cite{kim2010nanoscale,kim2011functional,lu2011two}. Subsequently we provide a Verilog-A based model.

A memristive device has a state variable $ \omega \subset [0,1] $ corresponding to the value of its memristance $ R_{ m} $. The $ R_{ m} $ is a function of $ \omega $, which follows
\begin{equation}\label{equ:Rm}
R_{ m} = \left\{ 
  \begin{array}{l l}
    R_{\rm OFF}(R_{\rm ON}/R_{\rm OFF})^\omega, & \quad (v\ge 0)\\
    R_{\rm OFF} ,& \quad (v < 0)
  \end{array} \right.
\end{equation}
where $ v $ is the applied voltage. According to Eq.~\ref{equ:Rm}, $ R_{ m} = R_{\rm OFF} $ when $ \omega = 0 $, while $ R_{ m} $ is in $ R_{\rm ON} $ state if $ \omega = 1 $.

The memristive device is observed with a well-defined threshold voltage. On one hand, if the absolute biased voltage applied to the memristive device is smaller than the threshold voltage, the state variable stays unchanged or changes minimally. On the other hand, once the bias voltage is larger than the threshold voltage, the state variable changes abruptly. Here we use symmetric threshold voltages, which can be different in practical memristive device implementations. The dynamic switching behavior of the memristive device is defined as:
\begin{equation}
{dw\over dt} = \left\{ 
  \begin{array}{l l l}
    \alpha (v - V_{\rm TH}), & \quad (v\ge V_{\rm TH})\\
    \alpha (v + V_{\rm TH}), & \quad (v\le -V_{\rm TH})\\
    \beta v, & \quad \rm otherwise,
  \end{array} \right.
\end{equation} 
where $ V_{\rm TH} $ is the threshold voltage. The $ \alpha $ and $ \beta $ coefficients are programing rates. Here, we set $ \beta = 0 $ assuming that the smaller voltage does not alter the state variable, and hence, does not perturb the memristance during read operation~\cite{lehtonen2014cellular}.

\begin{figure}
\centering
\includegraphics[trim=0 0 0 0,clip,width=0.45\textwidth]{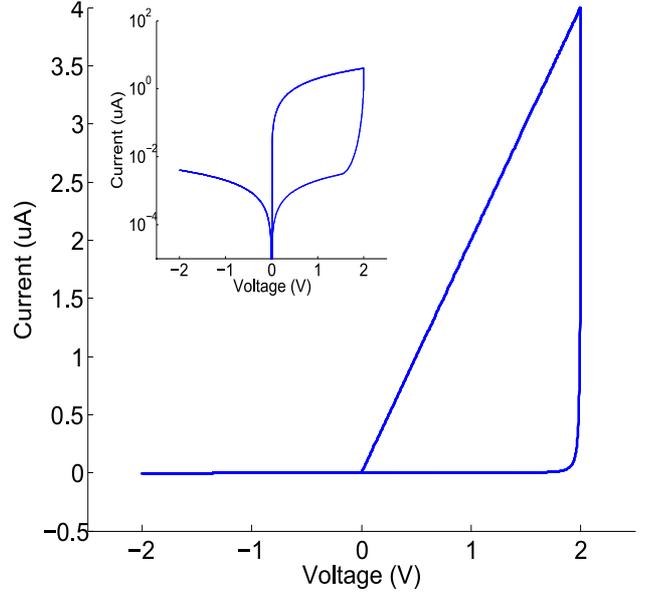}
\caption{Simulated I-V characteristic of the memristive device behavioral model with intrinsic diode-like behavior. It is obtained by using a sinusoid signal with 10~MHz frequency and 2.0 V (peak-to-peak voltage) amplitude. The inset figure illustrates the same I-V characteristic but the y-axis is on a logarithmic scale.}
\label{MemristorIV}
\end{figure}

The developed behavioral model is created in Verilog-A language is given in Appendix. Simulated I-V characteristic curve with intrinsic rectifying behavior is shown in Fig.~\ref{MemristorIV}.

\section{Results and Analyses}\label{ResultsAndAnalysis}
In this section, the aforementioned memristive device model is placed into crossbar array to study the read margin and power consumption under different crossbar sizes, read-schemes, interconnect resistance values, rectification ratios (assuming a fixed $ R_{\rm OFF}/R_{\rm ON}$ ratio for the defined memristive device model) and $ R_{\rm ON} $ value when the rectification ratio is constant.  

Simulations were carried out using Cadence tools, then data is extracted for post-processing in MATLAB. In these simulations the interconnect resistance is taken into consideration to produce a more realistic assessment of the crossbar array. The default parameters used in following simulations are listed in the Table~\ref{DefauParam}. Here, we mainly focus on the reading rather than the writing operation performance. For the following simulation we assume no change to the device memristance when a reading voltage in the order of 1 V is applied to the word/bit-lines when assuming a threshold voltage of 1.5 V.

\begin{table}[h]
\caption{List of parameters used in the memristive device and array simulations.}
\label{DefauParam}
\begin{center}
\begin{tabular}{ l  l }
\hline \hline
 High resistance state $ R_{\rm OFF} $ & $ 5\times 10^8~\Omega  $ \\
  Low resistance state $ R_{\rm ON} $ & $ 5\times 10^5~\Omega  $ \\
  Sense resistor $ R_{\rm sense} $ & $ 1.58\times 10^7~\Omega  $ \\
  Interconnect resistance $ r_{\rm wire}$ & $ 5~\Omega $\\
  Threshold voltage $ V_{\rm TH} $& $ 1.5  $~V\\
  Ratio of $ R_{\rm OFF}/R_{\rm ON} $ & $ 1\times 10^3 $\\
  $ \alpha $  & $2.5\times10^8$~(Vs)$^{-1}$\\ 
  $ \beta $&$0$~(Vs)$^{-1}$\\ \hline \hline
\end{tabular}
\end{center}
\end{table}

\subsection{Read-Schemes and Crossbar Size} \label{SizeAssess}
The read margin and power consumption under different crossbar sizes using different read-schemes are studied based on the device model developed earlier. The size of the crossbar is from $ 4\times 4 $ up to $ 128\times 128 $. The considered read-schemes are V/2, V/3 and F-F.

\begin{figure}[h]
\centering
\includegraphics[trim=0 0 0 0,clip,width=0.45\textwidth]{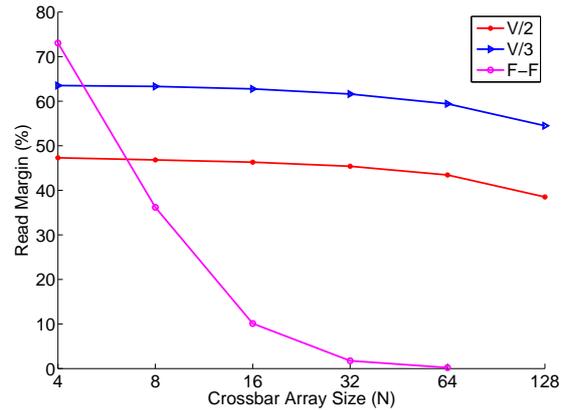}
\caption{Read margin as a function of crossbar size under different read-schemes. Where $ N $ is the number of columns or rows of crossbar. Here the number of columns equals the number of rows.}
\label{RMSize}
\end{figure}

\begin{figure}[h]
\centering
\includegraphics[trim=0 0 0 0,clip,width=0.45\textwidth]{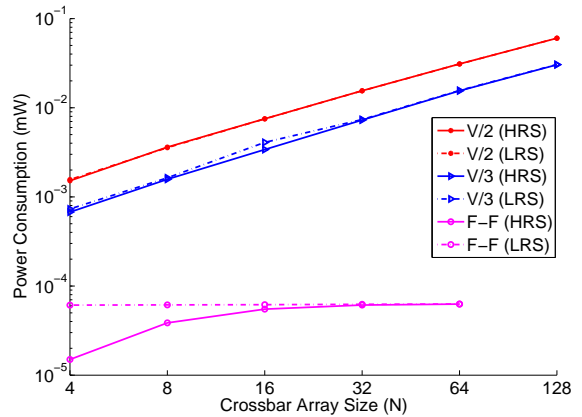}
\caption{Power consumption as a function of crossbar size under different read schemes. Where $ N $ is the number of columns or rows of crossbar.}
\label{PWSize}
\end{figure}

 Read operation performance results are shown in Figs~\ref{RMSize} $ \& $ \ref{PWSize}. Fig.~\ref{RMSize} illustrates that the V/3 read-scheme has the best read margin, while the F-F read-scheme has the worst read margin when the crossbar size increases to $ 8\times 8 $. The significant drop in read margin using the F-F read-scheme can be attributed to the exponential increase in the number of sneak current paths inside the crossbar array, despite the use of the intrinsic rectifying behavior. Note that the F-F read-scheme has the largest read margin when the crossbar size is small (here it is $ 4\times 4 $). In terms of the other two read-schemes, there is a slight decrease in the read margin for large array sizes.

Power consumption is presented in Fig.~\ref{PWSize}. The power consumed when the target memristive device stays either in the HRS or LRS is measured separately. It can be seen that the different power consumptions in the HRS and LRS is very small (almost the same), except for the F-F read-scheme when the crossbar size is small (smaller than $ 8\times 8 $ in our study). In overall, the F-F read-scheme consumes less power, whereas the V/2 read-scheme consumes the largest power. In particular power consumption when using F-F read-scheme is only a very small fraction of other two read-schemes, especially when using a large crossbar array size.

Based on these simulations the V/3 read-scheme has the best performance considering the trade-off between read margin and power consumption for large-scale crossbar arrays.

\subsection{Interconnect Resistance}
\begin{figure}[h]
\centering
\includegraphics[trim=0 0 0 0,clip,width=0.45\textwidth]{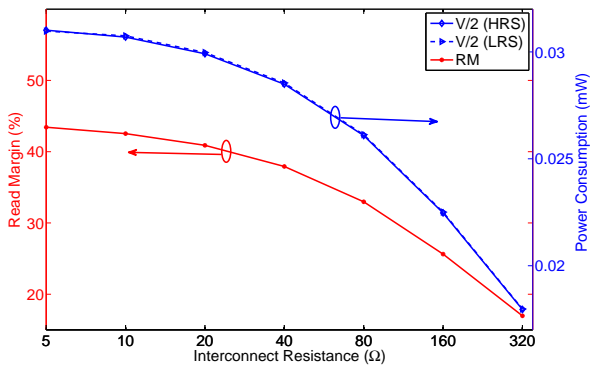}
\caption{ Read margin and power consumption influence as a function of interconnect resistance. The crossbar size is fixed at $ 64 \times 64 $. The V/2 read-scheme is used.}
\label{InterRes}
\end{figure}

The influence of interconnect resistance on read operation performance is shown in Fig.~\ref{InterRes}. As the interconnect resistance increases, the read margin decreases, while the power consumption reduces---in the other words, power consumption benefits from the increase in interconnect resistance. The read margin drops by 50\% as the interconnect resistance rises from $ 5~\Omega $ to $ 320~\Omega $. In order to achieve a high read margin performance, the interconnect resistance should be as small as possible despite the fact that a larger interconnect resistance can result in a slight improvement in the overall power consumption.    

Sources of read margin deterioration caused by interconnect resistance are: (i) voltage drops across the selected word/bit-lines; (ii) the unbalanced rising and falling voltage potential on unselected word/bit-lines, resulting in increasing many leakage current paths flowing into the selected bit-line. 
\subsection{Different Memristive Device Parameters}
\subsubsection{Resistance of ON/OFF}
\begin{figure}[h]
\centering
\includegraphics[trim=0 0 0 0,clip,width=0.45\textwidth]{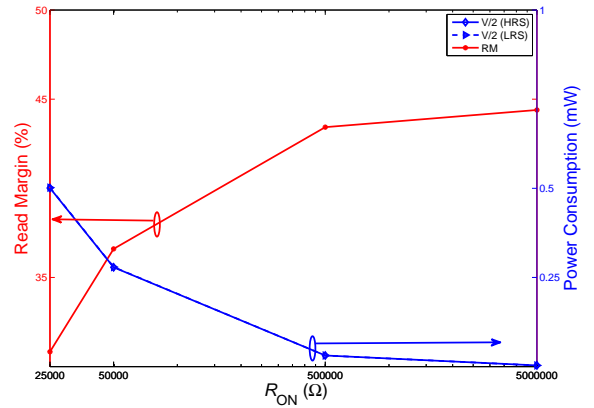}
\caption{ Read margin and power consumption as a function of $ R_{\rm ON} $ resistance. The crossbar size is fixed at $ 64 \times 64 $. Rectification ratio is a constant value, $ 10^3 $. The V/2 read-scheme is used.}
\label{LRSResAssess}
\end{figure}

The Fig.~\ref{LRSResAssess} shows the influence of the ON/OFF resistance on read operation. To ensure the ratio of HRS/LRS is kept constant during the simulation, the $ R_{\rm OFF} $ value is changed accordingly. As a result, the sensing resistor, $ R_{\rm sense} $ also needs to be changed to meet the requirement set by Eq.~\ref{Rsense}. 

The read margin improves if the $ R_{\rm ON} $ resistance increases.  The overall power consumption also improves as the $ R_{\rm ON} $ resistance increases. It appears that increasing $ R_{\rm ON} $ resistance produces improved performance with respect to both read margin and power consumption. However, it should be noted that there are still trade-offs among read margin, power consumption and read speed during read operation. For fast access times into memristive device based crossbar memory, higher resistance of the crosspoint cell always aggravates access speed.  

\subsubsection{Rectification Ratio Dependence}\label{RectRatio}
\begin{figure}[h]
\centering
\includegraphics[trim=0 0 0 0,clip,width=0.45\textwidth]{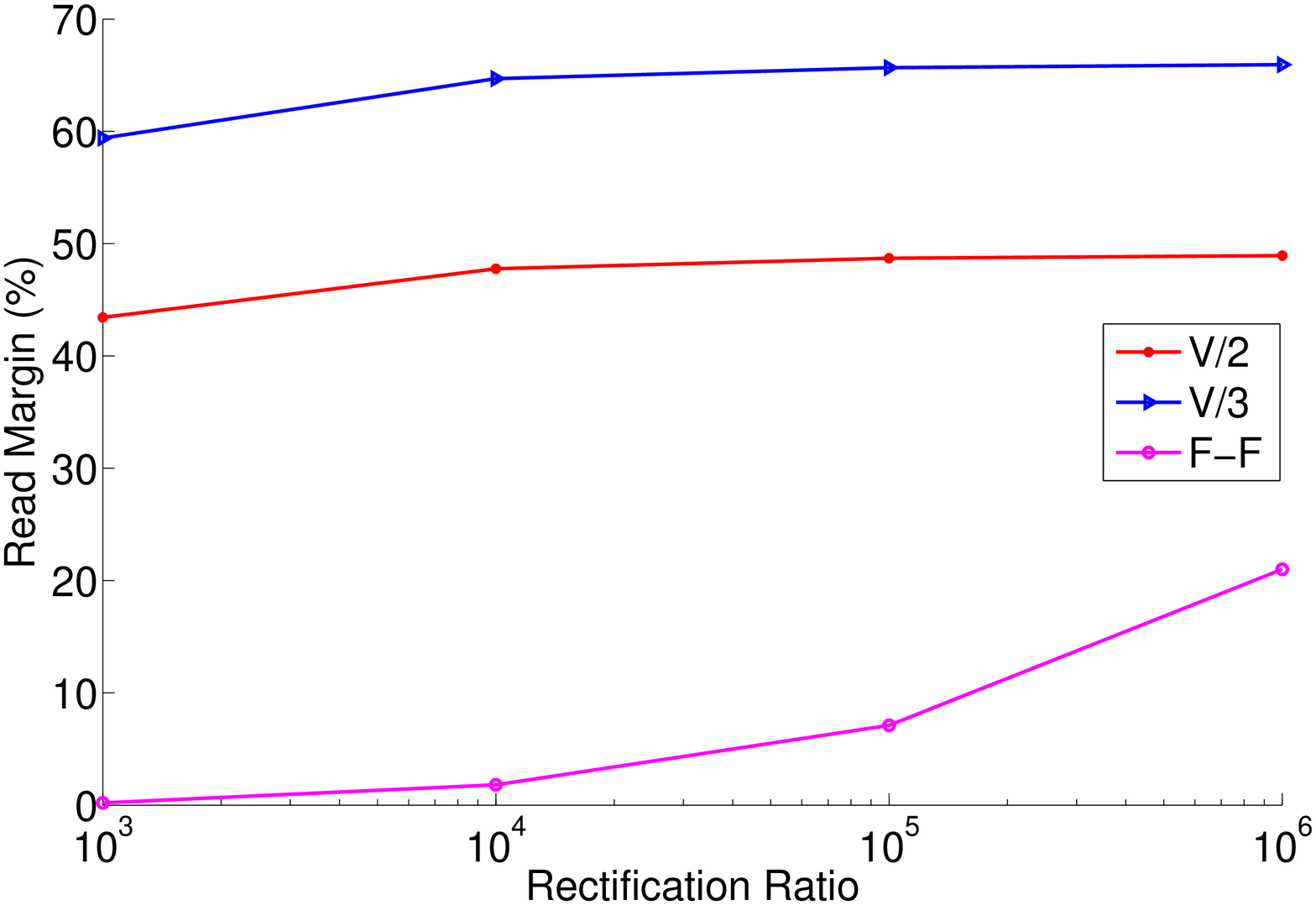}
\caption{Read margin as a function of different rectification ratios under different read-schemes.  The crossbar size is fixed at $ 64 \times 64 $. $ R_{\rm ON} $ is fixed at $ 5\times 10^5~\Omega $.}
\label{RMLRSFixRatioAssess}
\end{figure}

\begin{figure}[h]
\centering
\includegraphics[trim=0 0 0 0,clip,width=0.45\textwidth]{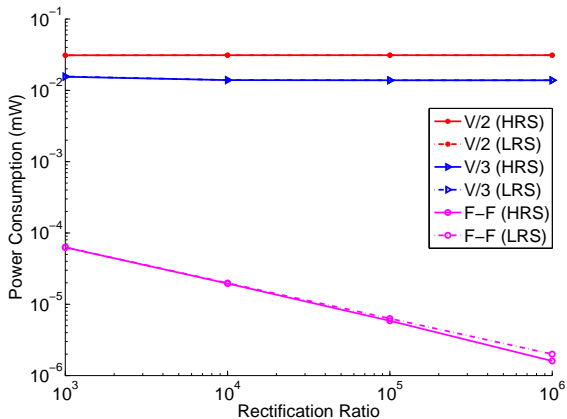}
\caption{Power consumption as a function of rectification ratios under different read-schemes. The crossbar size is fixed at $ 64 \times 64 $. $ R_{\rm ON} $ is fixed at $ 5\times 10^5\Omega $.}
\label{PWLRSFixRatioAssess}
\end{figure}

In this part, the $ R_{\rm ON} $ resistance remains unchanged, while the rectification ratio (here, it is also the ratio of HRS/LRS according to the mathematical model) changes to different values. The read margin and power consumption as functions of rectification ratios are shown in Figs.~\ref{RMLRSFixRatioAssess} $ \& $~\ref{PWLRSFixRatioAssess}, respectively.

It can be seen from Fig.~\ref{RMLRSFixRatioAssess} that read margin improves as the rectification ratio increases, especially for the the F-F read-scheme, which verifies that sneak currents can be suppressed significantly by memristive device with intrinsic rectifying behavior.

Power consumption almost stays constant for V/2 and V/3 read-schemes as the rectification ratio increases. While the power consumption with respect to F-F read-scheme reduces exponentially with increasing the rectification ratio.
%The reason is that most power consumption is consumed by the memristive devices in half-selected bit/word lines with respect to V/2 read scheme and one-third selected word line and two-third selected bit line with respect to V/3 read scheme. 
\section{Comparison and Discussion}\label{Comparison}
We have presented a systematic evaluations and analyses on the read operation performance of a crossbar array in Section~\ref{ResultsAndAnalysis}. In this section, we firstly compare our proposed model with one linear memristive device model. Secondly, we further compare the device model with the 1S1M structure. Intuitively one might suppose that increasing nonlinearity of the selector device in 1S1M structure is going to always improve the read operation performance of the 1S1M structure. However, the following simulations demonstrate this is not the case.

\begin{figure}[h]
\centering
\includegraphics[trim=0 0 0 0,clip,width=0.45\textwidth]{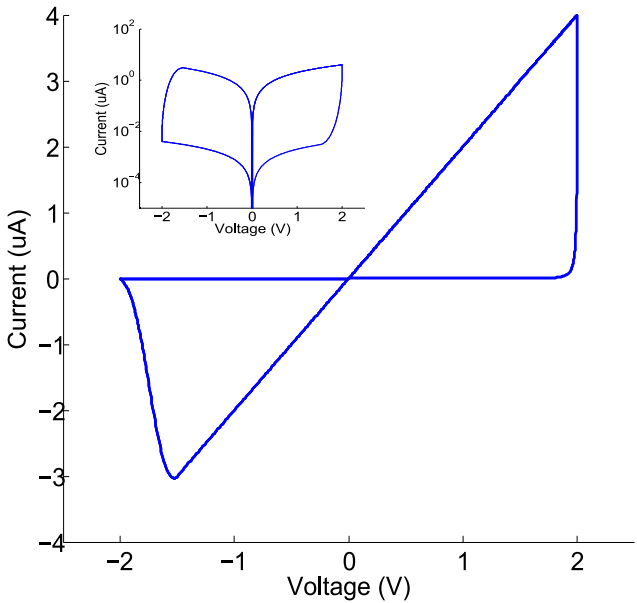}
\caption{Simulated I-V curve of the behavioral model without intrinsic rectifying characteristic. Simulation settings is same to Fig.~\ref{MemristorIV}. Inset figure shows the same I-V curve but the y-axis is on a logarithmic scale. }
\label{MemWithoutRect}
\end{figure}

\subsection{Comparison with the linear Memristive Device} \label{CompareLinearDevice}
\begin{figure}[h]
\centering
\includegraphics[trim=0 0 0 0,clip,width=0.45\textwidth]{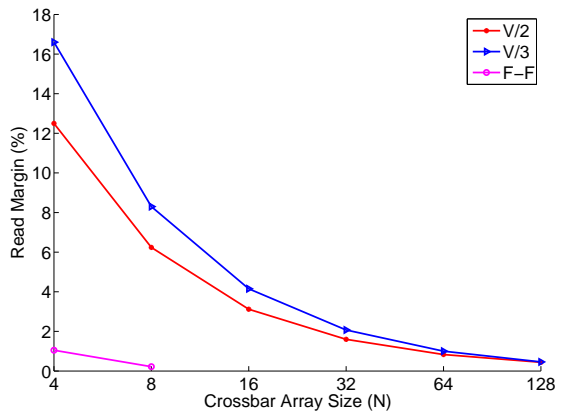}
\caption{Read margin as a function of crossbar size under different read-schemes. Where $ N $ is the number of columns or rows of crossbar.}
\label{RMSizeWithoutRect}
\end{figure}

\begin{figure}[h]
\centering
\includegraphics[trim=0 0 0 0,clip,width=0.45\textwidth]{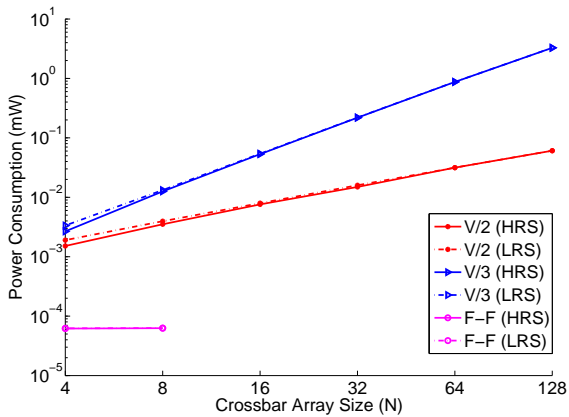}
\caption{Power consumption as a function of crossbar size under different read-schemes. Where $ N $ is the number of columns/rows of crossbar.}
\label{PWSizeWithoutRect}
\end{figure}

The I-V characteristic of linear memristive device model  (with linear resistance behavior in LRS State) is shown in Fig.~\ref{MemWithoutRect}, which has same parameter settings in Fig.~\ref{MemristorIV} except the rectifying behavior. We conduct the read margin and power consumption performance evaluation by using the linear memristive device model. The results are shown in Figs.~\ref{RMSizeWithoutRect} $\&$~\ref{PWSizeWithoutRect}. It can be seen that both of the read margin and power consumption performance deteriorate in comparison with the results in Figs.~\ref{RMSize} $ \& $~\ref{PWSize}. For read margin, the V/3 read-scheme still has the best performance and the F-F illustrates the worst read margin that is similar to results obtained in Section~\ref{SizeAssess}. Without self rectifying behavior, the sneak-path currents cannot be effectively suppressed, resulting in severely decreased read margin. Especially for the F-F read-scheme, the read margin is too small for practical application, even when considering small array size. Unlike the read margin performance, if the memristive device with self rectifying behavior employed does not see obvious deterioration as the size increases up to $ 128\times 128 $, the read margin employing linear memristive device suffers from a significant degradation  as the crossbar size increases.

%In contrast to the results presented in Section~\ref{SizeAssess}, Fig.~\ref{PWSizeWithoutRect}  where the V/3 read-scheme presented the best performance in terms of read margin and power consumption, when using a linear memristor model , this read-scheme results in a worse performance in both cases when increasing the crossbar array size.

The V/3 read-scheme has a worse power consumption than the V/2 read-scheme, which does not follow the results presented in Section~\ref{SizeAssess}, where the V/3 read-scheme has better power performance than the V/2 read-scheme. In addition, the other difference between the results in Section~\ref{SizeAssess} and the results in Fig.~\ref{PWSizeWithoutRect} is that the power consumption increases faster as the size of crossbar increases when using a linear memristive device.

\subsection{Comparison with the 1S1M Structure}\label{Compare1S1R}
\begin{figure}[h]
 \centering
 \includegraphics[trim=0 0 0 0,clip,width=0.45\textwidth]{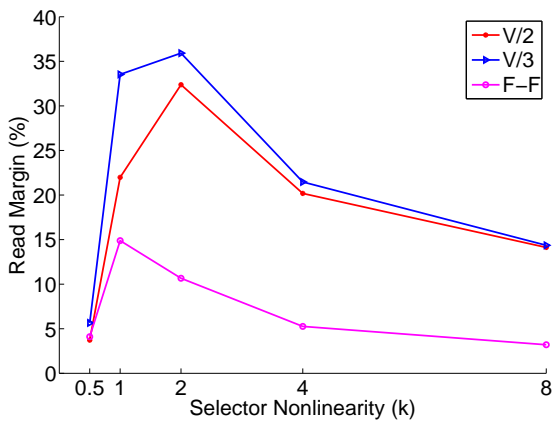}
 \caption{Read margin as a function of nonlinearity ($ k $) of the selector. The size of crossbar array is fixed at $ 64\times 64 $.}
 \label{RM1S1RNonlinearityAssess}
 \end{figure}

\begin{figure}[h]
 \centering
 \includegraphics[trim=0 0 0 0,clip,width=0.45\textwidth]{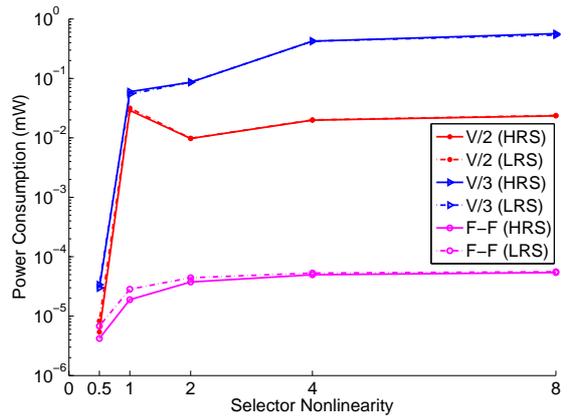}
 \caption{Power consumption as a function of nonlinearity ($ k $) of the selector. The size of crossbar array is fixed at $ 64\times 64 $.}
 \label{PW1S1RNonlinearityAssess}
 \end{figure}

In this part, we compare the read operation performance of the model presented in this paper with 1S1M structure. For 1S1M structure, the memory cell is made up of a selector and a memristive device. The selector has the characteristic of allowing both polarities to conduct beyond a threshold. The mathematical model is defined as \cite{zhou2014crossbar}:
  \begin{equation}
  I_{\rm sel} = {\gamma \times {\rm sinh}(\alpha \times V)},~~{\rm where}~~\alpha =  k\times p,
  \end{equation}
 the $ \gamma $ is a conductance parameter and $ k $ determines the nonlinearity of the selector. here, the $p = 18.4 $ is set to the same value presented in \cite{zhou2014crossbar}. To model the 1S1R structure, we connect one selector with one resistor acting as a memory cell. Here, $ R_{\rm ON}$ and $ R_{\rm OFF}$ are set to $5\times 10^5 $ and $5\times 10^8 $, respectively,  the rest of the parameters are the same as given in Table~\ref{DefauParam}. During these simulations, the value of  $ \gamma $ is kept fixed, while $ k $ was changed to determine the nonlinearity of the selector, which influences the read operation performance. Results are shown in Figs.~\ref{RM1S1RNonlinearityAssess} $ \& $~\ref{PW1S1RNonlinearityAssess}. It can be seen that there exists an optimum nonlinearity for the selector to achieve the best read margin, in contrast to the results when the rectification ratio is increase, which always results in an increased read margin as discussed in Section \ref{RectRatio}. So using the 1S1M structure, there is an issue that the nonlinearity should be cautiously considered along with the forward current $ I_{\rm ON} $ to acquire the maximum read margin. In addition, the power consumption always increases regarding for the 1S1M structure when the $ k $ increases, which is also different from the power consumption results reported in Section \ref{RectRatio} that which is not affected by the increase of rectification ratio, moreover, the F-F read-scheme does benefit from the rise of rectification ratio.
 
\section{Conclusion}\label{Conclu}
In this paper, the read operations of the crossbar array are studied with the proposed memristive behavioral model. Based on extensive simulation results, we verify that rectification behavior significantly improve the read margin. In addition, we provide circuit design guidelines that the V/3 read-scheme shows the best read margin and demonstrated that the F-F read-scheme presents the worst read margin as the crossbar size increases. However, the F-F read-scheme has the best power consumption performance due to a very small fraction of power consumption in comparison with the power consumed by the other two read-schemes (V/2 and V/3 read-schemes). Moreover, the interconnect resistance influence on read performance is also studied, which suggests that the resistance should be kept as small as enough to achieve better read margin. Furthermore, read operation dependence on memristive device parameters is investigated. Based on the numerical results, higher HRS/LRS resistance~--if the read speed has met with the requirement---and higher rectification ratio are desirable. Finally, we demonstrate the advantages of pursuing this intrinsic rectifying behavior within memristive device based on two comparisons in Section~\ref{Comparison}. From comparison in Section~\ref{CompareLinearDevice} we confirm that, as expected, the rectification does suppress sneak-path currents in crossbar memory leading to an improved figure-of-merit for the read operation. In terms of comparison in Section~\ref{Compare1S1R}, we show that increasing the nonlinearity of selector in 1S1M structure does not always lead to better read operation performance. In contrast, it can deteriorates the read margin and power consumption figure-of-merits without careful selection of this parameter. So the memristive device with intrinsic rectifying behavior is a significant improvement over alternative schemes for a crosspoint cell in the crossbar array memory. 
\section*{Acknowledgment}
This research was supported by a grant from the Australian Research Council (DP140103448). The authors would also like to thank the sponsorship from China Scholarship Council.
\section{Appendix}\label{Append}

\begin{lstlisting}{label}
`include "disciplines.vams"
`include "constants.h"
`include "constants.vams"

module memristor (p,n);
  	inout p;//positive terminal
  	inout n;//negative terminal
  	electrical p, n;
  	
	//parameter defination and default values
  	parameter real rof=5E8;                                 
  	parameter real ron = 5E5;
  	parameter real winit = 1;
  	parameter real Vth=1.5;
  	parameter real alpha=2.5E8;
  	parameter real beta=0;

  	real dwdt;
  	real w_normal_last;
  	real w_normal;
  	real R;
  	real first_iteration;
  	real stop;
	///////////////main/////////////////
	analog begin
		if (first_iteration==0) begin
		    w_normal_last=winit; 
		    //if this is the first iteration, start with winit;
		end
	
		if (V(p,n) >= Vth) begin
		    dwdt=alpha*(V(p,n)-Vth);
		end else if (V(p,n) <= -Vth) begin
		    dwdt=alpha*(V(p,n)+Vth);
		end else begin
		    dwdt=beta*V(p,n);
		end
	
		w_normal=idt(dwdt,w_normal_last,stop);	
		if ((w_normal>=1)&&(V(p,n)>0)) begin
		w_normal_last=1;
		stop=1;
		end
		else if ((w_normal<=0)&&(V(p,n)<0)) begin
		w_normal_last=0;
		stop=1;
		end
		else if (stop!=0) begin
		w_normal_last=w_normal;
		stop=0;
		end
		
		if((V(p,n)< 0)) begin
		R=roff;
		end 
		else if (V(p,n)>= 0) begin
		R=roff*pow(ron/roff,w_normal);
		end
	
		I(p,n) <+ V(p,n)/R;
		first_iteration=1;
	end
endmodule
\end{lstlisting}

\section*{References}

%\bibliography{mybibfile}

\end{document}